\documentclass[preprint,showpacs,preprintnumbers,amsmath,amssymb]{revtex4}

\usepackage{graphicx,setspace}% Include figure files
\usepackage{dcolumn}% Align table columns on decimal point
\usepackage{bm}% bold math
\usepackage{gensymb}
\usepackage{mathtools}
\usepackage{float}
\usepackage{soul}
\usepackage[title]{appendix}
\usepackage{xcolor}
\usepackage[per-mode=symbol]{siunitx}
\usepackage{amsmath}%nice math
\usepackage{mathrsfs} %robert smith formal script \mathscr{}
\usepackage{siunitx} % Required for alignment
\newcommand{\figref}[1]{Fig. \ref{#1}}
\newcommand{\authorcontributions}[1]{\section*{Author Contributions} \noindent #1}
\newcommand{\appref}[1]{\textit{Appendix \ref{#1}}}
\renewcommand{\eqref}[1]{Eq. \ref{#1}}
\usepackage{wrapfig}

\begin{document}

\preprint{}

% Still not convinced of this title
\title{Using colloidal deposition to mobilize immiscible fluids\\ from porous media}% Force line breaks with \\

%I don't know how to deal with multiple affiliations...a problem for later
% I just kind of cobbled together a hack that appears to work
\author{Joanna Schneider}
  \altaffiliation{Chemical and Biological Engineering, Princeton University, Princeton, NJ 08544\\$^\dag$To whom correspondence should be addressed: \texttt{ssdatta@princeton.edu}.}
\author{Rodney D. Priestley}
  \altaffiliation{Chemical and Biological Engineering, Princeton University, Princeton, NJ 08544\\$^\dag$To whom correspondence should be addressed: \texttt{ssdatta@princeton.edu}.}  \author{Sujit S. Datta$^\dag$}
  \altaffiliation{Chemical and Biological Engineering, Princeton University, Princeton, NJ 08544\\$^\dag$To whom correspondence should be addressed: \texttt{ssdatta@princeton.edu}.}  
\date{\today}

\begin{abstract}
Colloidal particles hold promise for mobilizing and removing trapped immiscible fluids from porous media, with implications for key energy and water applications. Most studies focus on accomplishing this goal using particles that can localize at the immiscible fluid interface. Therefore, researchers typically seek to optimize the surface activity of particles, as well as their ability to freely move through a pore space with minimal deposition onto the surrounding solid matrix. Here, we demonstrate that deposition can, surprisingly, \textit{promote} mobilization of a trapped fluid from a porous medium without requiring any surface activity. Using confocal microscopy, we directly visualize both colloidal particles and trapped immiscible fluid within a transparent, three-dimensional (3D) porous medium. We find that as non-surface active particles deposit on the solid matrix, increasing amounts of trapped fluid become mobilized. We unravel the underlying physics by analyzing the extent of deposition, as well as the geometry of trapped fluid droplets, at the pore scale: deposition increases the viscous stresses on trapped droplets, overcoming the influence of capillarity that keeps them trapped. Given an initial distribution of trapped fluid, this analysis enables us to predict the extent of fluid mobilized through colloidal deposition. Taken together, our work reveals a new way by which colloids can be harnessed to mobilize trapped fluid from a porous medium.
% Surface-active colloidal particles can reduce the interfacial tension between two immiscible fluids; thus, they hold promise for mobilization of trapped immiscible fluids from porous media. To accomplish this, researchers typically focus on optimizing the surface activity of particles, as well as their transport through a pore space with minimal deposition onto the surrounding solid matrix, which is thought to hinder subsequent flow. Here, we demonstrate that deposition can, surprisingly, \textit{improve} mobilization of a trapped fluid from a porous medium without requiring any surface activity. Using confocal microscopy, we directly visualize both colloidal particles and trapped immiscible fluid within a transparent, three-dimensional (3D) porous medium. We find that as non-surface active particles deposit on the solid matrix, increasing amounts of trapped fluid become mobilized. Furthermore, by directly visualizing the extent of deposition and the geometry of trapped fluid droplets at the pore scale, we quantify how deposition increases the viscous forces on trapped droplets, overcoming the capillary forces that keep them trapped and enabling them to be mobilized. Together, our work suggests a new way by which colloidal deposition can be harnessed to mobilize trapped fluid from a porous medium, with implications for key energy and water applications.

\end{abstract}

%PACS, the Physics and Astronomy Classification Scheme
\pacs{Valid PACS appear here}

% Keywords
% \keywords{}
%Use showkeys class option if keyword display desired

\maketitle
\bibliographystyle{ieeetr}

\section{\label{sec:level1}Introduction}

\noindent Studies of colloidal particles in porous media inform their use in broad applications \cite{Bizmark2019} including groundwater remediation \cite{Phenrat2007,Phenrat2011} and enhanced oil recovery (EOR) \cite{Hendraningrat2013,Hussain2013,Zhang2014}. The ability to predict whether and how nano- and micro-scale particles can help mobilize trapped immiscible fluids from a heterogeneous pore space is critically important in both of these applications. For example, non-aqueous contaminants in groundwater aquifers pose a major risk to human health \cite{McKnight2010,Russell1992}; the widespread use of organic compounds as industrial solvents along with improper disposal have caused them to infiltrate aquifers, where they remain trapped by capillarity \cite{Mackay1985}. Thus, remediation efforts are exploring the use of colloidal particles to help mobilize these contaminants \cite{Corapcioglu1993,Roy1997,DeJonge2004}. Another example is that of oil recovery: while primary and secondary recovery processes are commonly employed, they still leave up to $90\%$ of the oil in a subsurface reservoir behind, again due to trapping by capillarity \cite{Wardlaw1979}. As global energy demand rises, researchers are therefore increasingly exploring the use of colloidal particles to help mobilize trapped oil for enhanced recovery \cite{Suleimanov2011,Roustaei2015,Wei2016,Li2017}.

% Hence, colloidal particles are frequently designed or chosen specifically to maximize their surface activity.

One mechanism by which colloidal particles are thought to mobilize trapped immiscible fluid is through surface activity. In this mechanism, particles preferentially localize at immiscible fluid-fluid interfaces owing to their surface chemistry, thereby reducing the interfacial tension and weakening the capillary stresses that keep fluid trapped \cite{Roy1997,Franzetti2009,Franzetti2010,Hendraningrat2013,Alomair2014,Ragab2015,Magro2016}. Non-surface active colloids also have potential for immiscible fluid mobilization---for example, through their ability to reduce fluid slip \cite{Yu2015} or enhance disjoining pressure \textit{via} the formation of thin films \cite{Kondiparty2011,Zhang2014}. Importantly, all of these mechanisms rely on the ability of colloidal particles to freely move through the heterogeneous pore space without depositing on the solid matrix: deposition hinders the ability of particles to localize at target fluid interfaces, \cite{Kuhnen2000,Phenrat2007} and potentially causes clogging that impedes subsequent flow \cite{Wiesner1996,Chen2008,Civan2010,Hendraningrat2013a}. As a result, researchers typically seek to minimize deposition of colloidal particles in porous media---for example, by tuning colloidal interactions or only using low concentrations of particles \cite{Bradford2008,Zhang2010}.

Here, we demonstrate that deposition can, surprisingly, \textit{promote} mobilization of a trapped immiscible fluid from a 3D porous medium, without requiring surface activity. We do this by using refractive index-matching to render the medium transparent, which enables us to simultaneously visualize colloidal deposition and immiscible fluid droplets within the pore space. As the particles deposit on the solid matrix under an imposed flow rate, increasing amounts of trapped fluid become mobilized, ultimately enabling removal of $\sim70\%$ of the fluid. By analyzing the extent of deposition, as well as the geometry of trapped fluid droplets at the pore scale, we determine that deposition reduces the permeability of the medium---thereby enabling the viscous stresses on the trapped droplets to overcome the influence of capillarity keeping them trapped without excessively restricting the fluid flow. We develop a geometric model that quantifies changes in the permeability due to particle deposition, providing a first step toward predicting the extent of fluid mobilization given any initial distribution of trapped fluid. Our work reveals that colloidal deposition can be harnessed to mobilize trapped immiscible fluids and provides quantitative guidelines for applying this new mechanism.

\section{Establishment of an initial trapped fluid configuration}

\noindent We prepare a model 3D porous medium by lightly sintering a dense packing of borosilicate glass beads at 1,000 \degree C for 3.5 \si{\minute} in a quartz capillary tube with a square cross-sectional area $A = 9$ \si{\milli\meter}$^2$. The beads have radii $a = 62$--$75$ \si{\micro\meter} and the packing spans a length $\ell=2.6$ \si{\centi\meter}. Light scattering from the solid matrix formed by the beads typically renders the medium opaque, precluding imaging of flow and transport in the pore space. To overcome this limitation, we formulate a wetting fluid of 50 wt\% dimethyl sulfoxide (DMSO), 30 wt\% glycerol, and 20 wt\% deionized water, dyed with Rhodamine Red, that has the same refractive index as the solid matrix; thus, infiltrating the pore space with this fluid renders the medium completely transparent. Further, this fluid enables the colloidal particles used in our experiment to remain stable in suspension over an experimental time scale of $>18$ hours, as verified by confocal microscopy at single-particle resolution. To investigate immiscible fluid trapping and mobilization, we formulate another fluid, an oily mixture of aliphatic and aromatic hydrocarbons (\textit{Cargille} refractive index liquids), that has the same refractive index as the wetting fluid and the solid matrix, but is non-wetting. Measurements performed on a similar fluid pair indicate that the three-phase contact angle between the wetting fluid and glass in the presence of the non-wetting oil is $\theta\approx5$ \degree \cite{Krummel2013}. The interfacial tension between the two liquids is $\gamma \approx 13$ \si{\milli\newton}/\si{\meter} as measured previously \cite{Datta2014} and the dynamic shear viscosities of the wetting and non-wetting fluids are $\mu_w = 10.8$ \si{\milli\pascal}$\cdot$\si{\second} and $\mu_{nw} = 16.8$ \si{\milli\pascal}$\cdot$\si{\second}, respectively, measured using a cone-plate rheometer. Before saturating the pore space with fluid, we pull vacuum and push CO$_2$ through the medium. We then saturate the pore space with deionized water to dissolve any remaining CO$_2$. Finally, we pull vacuum once more before saturating the pore space with the dyed particle-free wetting fluid---preventing trapping of any residual air bubbles in the medium.

\begin{figure}[htp]
\centering
\includegraphics[width=0.87\textwidth]{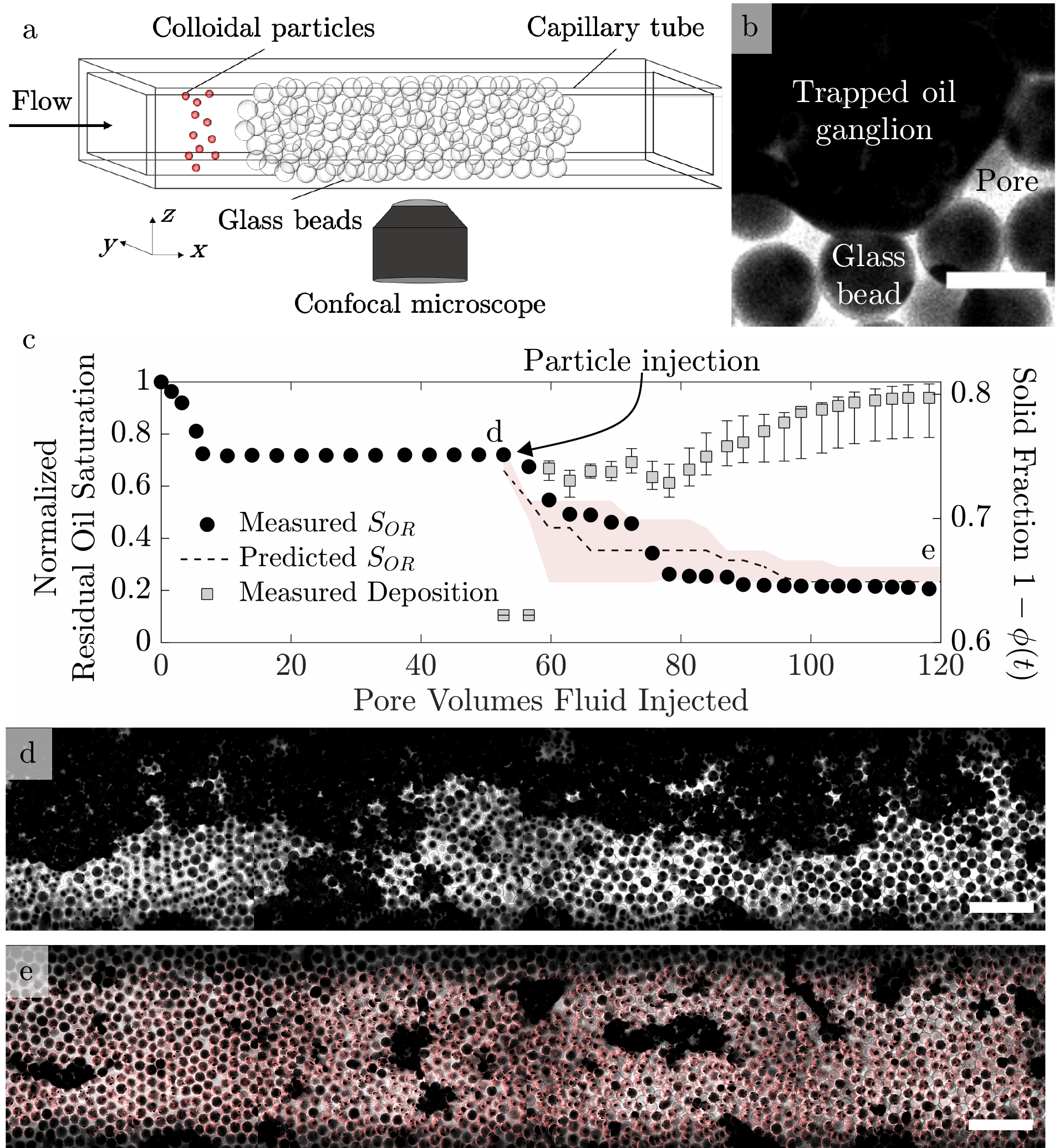} 
\caption{Colloidal deposition promotes oil mobilization. (a) Schematic of the porous medium composed of glass beads packed in a capillary tube. Fluid flow is imposed along $+x$ and micrographs are taken at a fixed depth $z$ within the medium. (b) Magnified micrograph showing trapped oil before particles are injected. Fluorescent signal from the dyed wetting fluid appears as white; hence, white indicates pore space, black circles show the beads, and additional black shows a trapped oil ganglion. Scale bar represents 150 \si{\micro\meter}. (c) Black points show the variation of the residual oil saturation $S_{OR}$, normalized by its maximal value 0.75, over the course of the experiment. Injecting particle-free wetting fluid results in an initial slight decrease in $S_{OR}$. We then inject the particle suspension at the time indicted by the arrow, which results in a further decrease in $S_{OR}$. Grey points show the increase in the solid fraction $1-\phi(t)$ over time due to deposition measured from the micrographs. Error bars show differences in the calculated solid area fraction for $\pm 20\%$ of the image binarization threshold. Dashed line shows our theoretical prediction of $S_{OR}$, and the red region shows uncertainty in the prediction arising from heterogeneities in deposition as described in Fig. 4. Panels (d) and (e) show images of the porous medium before and after colloidal injection, respectively, as indicated in (c), showing that particle deposition---indicated by red---mobilizes trapped oil. Scale bars represent 750 \si{\micro\meter}. Imposed flow direction is from left to right.}
\label{Figure 1}
\end{figure}

We then use a \textit{Nikon} A1R+ confocal fluorescence microscope to characterize the pore space structure, as schematized in \figref{Figure 1}a. Specifically, we obtain images, each spanning 1,024 \si{\micro\meter} $\times$ 1,024 \si{\micro\meter} in the $xy$ plane at a single depth within the porous medium centered at $z=135$ \si{\micro\meter}, with an optical thickness of 14 \si{\micro\meter}. A magnified view from one image is shown in \figref{Figure 1}b. The brightness in the image reflects the fluorescent signal from the dyed wetting fluid in the pore space, with primary excitation and emission at $560$ and $580$ \si{\nano\meter} respectively, enabling identification of the glass beads by their contrast with the dyed wetting fluid and shown by the dark circles in the bottom of the image. Together, the tiled images span the entire cross-section of the medium and therefore reproduce a two-dimensional (2D) slice of the pore space. Using adaptive binarization of the images and restricting all analysis to $\approx5$ bead diameters away from the inlet and outlet to minimize the influence of boundaries, we measure the initial porosity of the pristine medium, $\phi_0 = 0.38\pm0.06$, in good agreement with previous measurements \cite{Krummel2013}.

We introduce the non-wetting oil into the pore space using a \textit{Harvard Apparatus} Pump 11 Elite syringe pump, injecting $\approx$ 30 pore volumes (PVs) at a fixed volumetric flow rate $Q_{nw}=15$ \si{\milli\liter}/\si{\hour}; here, the number of PVs injected after a duration $t$ is given by $Q_{nw}t/\left(\phi_0 A\ell\right)$. We simultaneously acquire tiled images spanning the entire cross-section of the medium successively. Because the oil is undyed, it appears dark in the confocal micrographs, as indicated in the top of \figref{Figure 1}b. Thus, comparing the micrographs to those obtained during initial characterization of the pore structure provides a 2D map of the oil configurations at sub-pore resolution. Due to the high flow rate $Q_{nw}$, oil saturates the majority of the pore space. The measured residual oil saturation $S_{OR}$, defined as the fraction of the pore space area occupied by oil, is $0.75$. We normalize all subsequent measurements of $S_{OR}$ by this maximal value, as indicated by the first data point in \figref{Figure 1}c.

We then re-inject $\approx$ 55 PVs of the particle-free wetting fluid at a rate of $Q_w = 0.15$ \si{\milli\liter}/\si{\hour}, emulating flow conditions in many subsurface formations \cite{Medici2019}; in this case, $\text{PV}\equiv \int_0^tQ/\left(A\ell\phi(t)\right)\mathrm{d}t$ and the time-dependent porosity $\phi(t)$ measured directly from the confocal micrographs accounts for the fraction of the pore space accessible to the wetting fluid, which changes as oil is mobilized from the porous medium. Under these conditions, the wetting fluid flow leads to the formation of discrete droplets, or ganglia, of the non-wetting oil. While some of these ganglia become mobilized by the wetting fluid, $\approx72\%$ of the original oil in place remains trapped by capillarity \cite{Payatakes1984,Datta2014}, as indicated by the decrease in $S_{OR}$, shown by the black circles in \figref{Figure 1}c, within the first $\approx6$ PVs of fluid injection. The wetting fluid then continues to flow around them. This protocol thus establishes an initial steady-state configuration of trapped ganglia, as indicated by the circles for $\approx6$--$50$ PVs in \figref{Figure 1}c and exemplified by the micrograph in \figref{Figure 1}d. As is frequently the case in confined and stratified media \cite{Walker1998,Lashanizadegan2007}, most of the trapped ganglia are located near the boundaries of the medium, with smaller ganglia closer to the center of the pore space.

\section{Mobilization of trapped oil by colloidal particles}

\noindent To explore whether and how colloidal particles can help mobilize the trapped oil, we next inject a suspension of positively charged, amine-terminated polystyrene particles (\textit{Sigma Aldrich}) of diameter $d_p=1$ \si{\micro\meter} into the medium. The particles are fluorescently labeled, with primary excitation and emission at $470$ and $505$ \si{\nano\meter} respectively; these values are distinct from those of the wetting fluid, enabling multiplexed imaging of both particles and wetting fluid simultaneously. The spatial resolution of the imaging is 3 \si{\micro\meter}, and thus, each pixel comprises up to 3 particles. We first homogenize a concentrated 2.5 vol\% aqueous suspension of the particles by vortexing for 10--15 min, and then redisperse the particles in the wetting fluid at a concentration of $8\times10^{-3}$ vol\%. We then inject this suspension into the porous medium directly following the particle-free wetting fluid injection, at the same flow rate $Q_w = 0.15$ \si{\milli\liter}/\si{\hour}. The initiation of particle injection is indicated by the arrow in \figref{Figure 1}c. As injection progresses, the particles gradually deposit onto the negatively-charged glass matrix throughout the medium due to electrostatic attraction \cite{Bizmark2020}, indicated by the red in \figref{Figure 1}e. The measured solid area fraction averaged over the entire medium, $1-\phi(t)$, increases due to deposition and eventually plateaus after $\approx40$ PVs of the suspension are injected, as shown by the grey squares in \figref{Figure 1}c, likely due to electrostatic repulsion between like-charged particles \cite{Bizmark2020}.

By constricting the pore space, this process of deposition is usually thought to detrimentally hinder fluid flow. Unexpectedly, however, we find that trapped ganglia are concomitantly mobilized from the medium: the residual oil saturation decreases as deposition progresses, ultimately reaching only $\approx30\%$ of its initial steady-state value after $\approx30$ PVs of suspension are injected, as shown by the black circles for $\approx50$--$80$ PVs in \figref{Figure 1}c and exemplified by the micrograph in \figref{Figure 1}e. Clearly, particle deposition \textit{promotes}, not hinders, trapped oil mobilization.

\section{Particle Deposition Characteristics}
\noindent Why does particle deposition promote oil mobilization? Two possible mechanisms are (i) colloidal surface activity, which reduces the fluid-fluid interfacial tension and thereby reduces the strength of capillarity keeping ganglia trapped, and (ii) selective clogging of some pores redirecting the fluid flow toward trapped ganglia. We use our pore-scale visualization to examine the influence of these mechanisms in our experiment. 

\begin{figure}[htp]
\centering
\includegraphics[width=\textwidth]{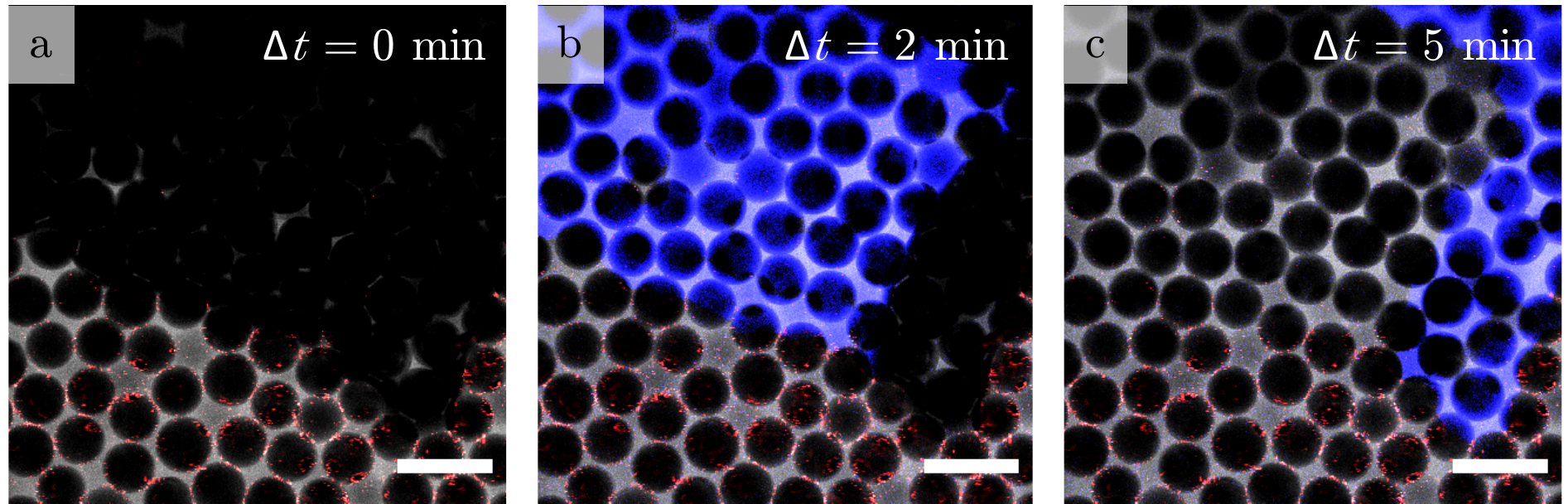}
\caption{Oil mobilization does not occur due to colloidal surface activity. Magnified views of confocal micrographs show an oil ganglion (a) before, (b) during, and (c) after mobilization. White regions show pore space, black circles show the bead matrix, additional black regions show oil, with the additional red showing colloidal particles. Some particles appear inside the ganglion due to the non-zero optical slice thickness. Particles are deposited on the bead surfaces, but do not localize at the ganglion surface, indicating that they are not surface active. Blue regions in (b) and (c) show portions of the ganglion that are mobilized between (a--b) and (b--c). Scale bars represent 250 \si{\micro\meter}. Imposed flow direction is from left to right.}
\label{Figure 2}
\end{figure}

We first consider the possibility that the particles are surface active and localize at the ganglia surfaces. Previous studies of polystyrene colloids indicate that amine-functionalized particles are only weakly surface active \cite{Kim2008,Feng2014}, suggesting that this mechanism for ganglion mobilization is unlikely. This suggestion is confirmed by direct inspection of the distribution of particles during ganglion mobilization. We do not observe particle localization at the ganglia surfaces, as exemplified in \figref{Figure 2}. The dark region in the top of \figref{Figure 2}a shows a portion of the ganglion immediately before it is mobilized; notably, the particles, shown by the red dots, deposit on the surrounding glass bead matrix, but do not noticeably localize at the ganglion surface. This feature is also apparent in \figref{Figure 2}b, which shows the ganglion in black as it is mobilized, with the portion of the ganglion that has been mobilized in this time step shown in blue; the boundary of this mobilized region is distinctly particle-free. We observe the same lack of particle localization in \figref{Figure 2}c, which shows the ganglion immediately after it has been mobilized from the field of view, with the remaining portion of the ganglion that has been mobilized shown in blue. The ganglion is mobilized as particles are deposited onto the surrounding bead matrix; however, we again do not observe any particle localization at the boundary of the mobilized region. Thus, mobilization of trapped oil is not due to colloidal surface activity.

 \begin{figure}[htp]
\centering
\includegraphics[width=\textwidth]{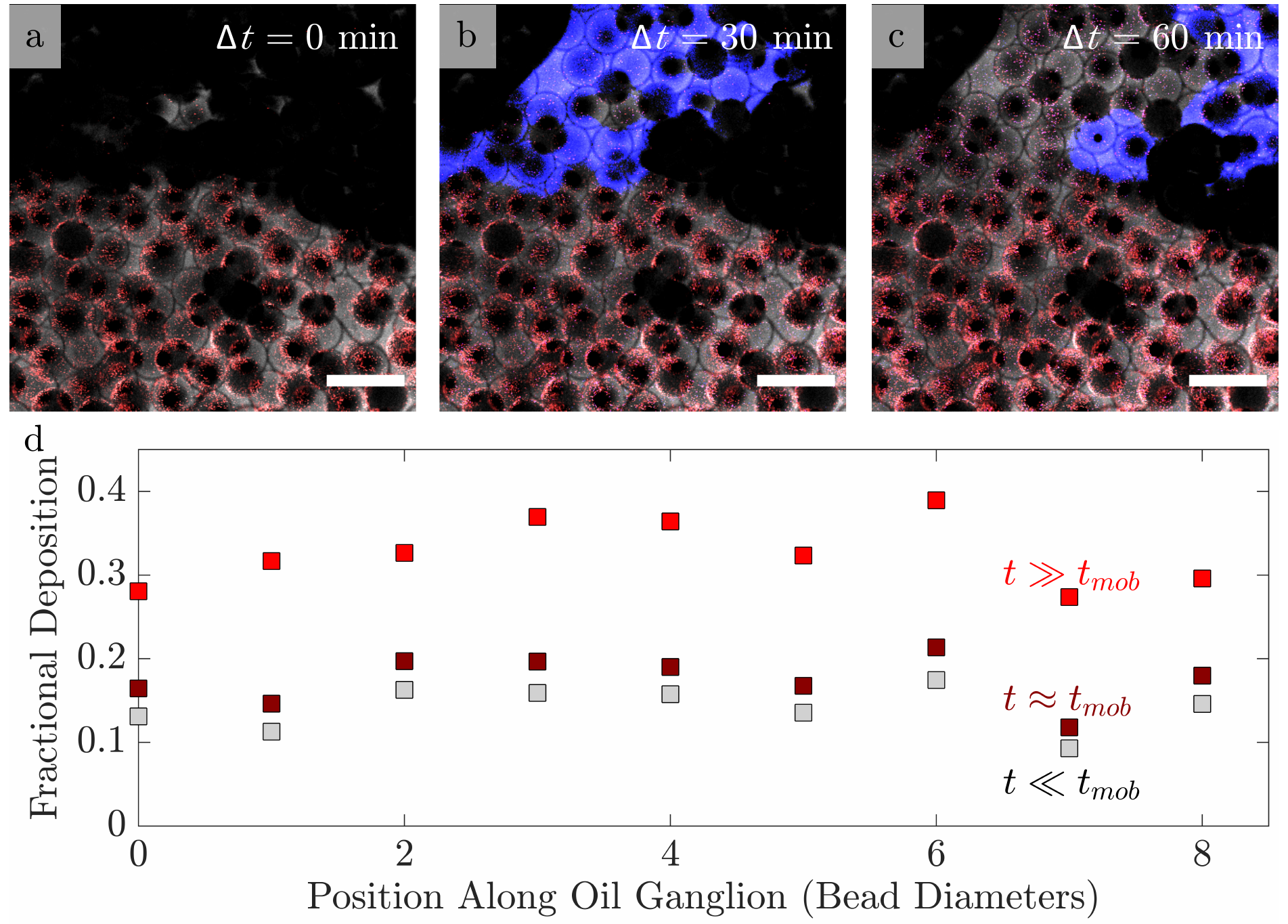}
\caption{Oil mobilization does not occur due to selective clogging of pores driving flow redirection. Magnified views of confocal micrographs show an oil ganglion (a) before, (b) during, and (c) after mobilization. White regions show pore space, black circles show the bead matrix, additional black regions show oil, with the additional red showing colloidal particles. Some particles appear inside the ganglion due to the non-zero optical slice thickness. The appearance of glass beads that are not completely black and have a shaded outline is a result of imperfect index matching between the wetting fluid and the beads. Particles are uniformly deposited on the bead surfaces both along and transverse to the ganglion; we do not observe selective clogging of pores. Blue regions in (b) and (c) show portions of the ganglion that are mobilized between (a--b) and (b--c). Scale bars represent 250 \si{\micro\meter}. Imposed flow direction is from left to right. (d) Data showing the fraction of the pore space immediately transverse to the ganglion that is occupied by deposited particles, measured using the confocal micrographs at three different times. Grey points correspond to (a), maroon points correspond to (b), and red points correspond to the end of the experiment, shown in Fig. 1e. In all cases, the deposition is uniform across pores.}
\label{Figure 3}
\end{figure}

    Next, we consider the possibility of flow redirection as a result of pore clogging. In particular, as particles deposit on the bead matrix, they may constrict and even occlude tight constrictions, possibly redirecting subsequent flow through the pore space. However, previous studies indicate that this behavior requires a considerable amount of deposition \cite{Wyss2006,Dersoir2015}, which is unlikely to occur given the dilute volume fraction of particles ($<10^{-2}$ vol\%) used in our experiment and the limited overall duration of particle injection. Indeed, as previously determined using experiments on single pores \cite{Wyss2006}, the total number of particles that must flow through each pore to clog it, $N^*$, can be estimated using the empirical scaling relation $N^*\approx5,000(2a_t/d_p)^4$, where $2a_t$ and $d_p$ are the diameter of a pore constriction (a ``throat") and a particle, respectively. In our experiment, $a_t\approx0.16a=11$ \si{\micro\meter} \cite{Al-Raoush2005,Thompson2008} and $d_p=1$ \si{\micro\meter}, yielding a threshold number of particles required to clog each pore, $N^*\approx1\times10^9$. By contrast, the characteristic number of particles that flow through each pore over the entire experimental duration---calculated using the imposed flow rate $Q=0.15$ mL/h, the maximal amount of time during which particles flow through the system $t\approx10$ h, the number of pore volumes of particle suspension flowed over that time PVs $\approx65$, the suspension volume fraction of $8\times10^{-3}$ vol\%, and the individual particle diameter of 1 \si{\micro\meter}---is $N_{tot}\approx4\times10^8$, much smaller than the estimated $N^*$. Moreover, the number of particles that flow through each pore before the onset of oil mobilization is even smaller than this value. Thus, clogging is not likely to occur in our experiment.

This expectation is again confirmed by direct inspection of the distribution of particles during ganglion mobilization: we do not observe pore clogging or strong variations in deposition in the different pores away from trapped ganglia, as exemplified by the micrograph shown in \figref{Figure 1}e, as well as the magnified views of ganglia mobilization shown in Figs. \ref{Figure 2} and \ref{Figure 3}. The dark region in the top of \figref{Figure 3}a again shows a portion of a ganglion immediately before it is mobilized; in this case, the ganglion is still trapped in the porous medium 2 h, or after 13 PVs of further fluid injection, after that shown in \figref{Figure 2} is mobilized, resulting in more deposition on the surrounding beads. We again observe little to no particle localization at the ganglion surface, consistent with the results shown in \figref{Figure 2}. Notably, we do not observe appreciable variations in particle deposition from pore to pore, either along the flow direction or laterally. Further, though they are increasingly constricted, the individual pores still permit fluid flow through them; particles only occupy $<20\%$ of the available pore space area, as shown by the grey points in \figref{Figure 3}d determined directly from the confocal micrographs. These features are also apparent in Figs. \ref{Figure 3}b--c, which again show mobilized portions of the ganglion in blue. After the ganglion at the top of the micrograph is mobilized, new particles are re-deposited in its wake, shown by the additional red in the upper region of \figref{Figure 3}c, helping to even out the spatial variations in permeability. Indeed, during and well after mobilization, we again find that particle deposition is nearly uniform across pores, as shown by the maroon and red points in \figref{Figure 3}d, consistent with previous findings \cite{Gerber2020}. Thus, mobilization of trapped oil is not due to selective clogging of pores. 

Our visualization of particle deposition reveals that it constricts but does not clog pores throughout the medium---and that the extent of oil mobilization is correlated with the extent of deposition. These observations hint at a different mechanism of oil mobilization. We hypothesize that deposition reduces the permeability of the medium, thereby enabling the viscous stresses exerted by the flowing wetting fluid on the trapped ganglia to overcome the influence of capillarity keeping ganglia trapped. To test this hypothesis, we use our micrographs to estimate the permeability reduction, thereby providing an estimate of how the local viscous stresses are modified, as detailed in the following section.

\section{Permeability reduction by colloidal deposition}
\noindent To accomplish this goal, we first establish the connection between the position- and time-dependent porosity of the medium, $\phi(x,t)$, and the thickness of the layer of particles deposited on the bead matrix, $\varepsilon(x,t)$; here, $x$ is the position along the imposed flow direction as indicated in \figref{Figure 1}a and both quantities are averaged laterally over the $y$ direction. In our experiment, we acquire data at a single $z$-location and assume that $\phi$ and $\varepsilon$ are uniform in the $z$-direction, since the medium is randomly packed and previous experiments and simulations have shown the flow to be statistically uniform throughout \cite{Kouri1996,Freund2003}.

Our pore-scale visualization of deposition shown in Figs. \ref{Figure 2}, \ref{Figure 3}, and \ref{Figure 4}a indicates that particles are deposited in thin, nearly uniform layers around the spherical glass beads. We rationalize this uniform pore-scale deposition profile by calculating the particle P\'{e}clet number, which describes the ratio of advective to diffusive transport, $\mathrm{Pe} \equiv \left(Q_w/A\right)/\left(\mathscr{D}/d_p\right)$, where $Q_w$ is the imposed flow rate, $A$ is the cross-sectional area of the porous medium, $\mathscr{D}$ is the particle diffusivity, and $d_p$ is the particle diameter. We calculate $\mathscr{D}$ using the Stokes-Einstein equation, $\mathscr{D} = k_BT/3\pi\mu_wd_p$, where $k_B$ is Boltzmann's constant and $T=300$ \si{\kelvin} is the temperature of the fluid, yielding $\mathrm{Pe} \approx 100$ for our experiment. While this calculation indicates that advective transport by the mean flow slightly dominates over diffusive transport of single particles, it does not fully capture the complicated flow in a porous medium. Previous measurements indicate that the flow velocities deviate strongly from the mean, in many cases being much smaller \cite{Datta2013}, particularly upstream of each bead \cite{Huang2008}, potentially leading to uniform deposition around the beads. This expectation has been confirmed in recent experiments investigating polystyrene particle transport in glass bead packings, which reveal that the interplay between particle interactions and flow limit deposition on the upstream faces of the beads and promote uniform deposition around each bead surface, even at $\text{Pe} \gtrsim 100$ \cite{Gerber2019}. Thus, for simplicity, we approximate deposition at the pore scale to be uniform around the individual beads, with a deposition thickness $\varepsilon$, as schematized in \figref{Figure 4}b. This thickness is then given by \begin{equation}
    \varepsilon(x,t)\approx a\left[\left(\frac{1-\phi(x,t)}{1-\phi_{0}}\right)^{1/3}-1 \right]
    \label{Porosity}
\end{equation}
% \begin{equation}
%     \phi(x,t) \approx 1-\left(1-\phi_0\right)\left(\frac{a+\varepsilon(x,t)}{a}\right)^3
%     \label{Porosity}
% \end{equation}
where $\phi_{0}$ is the initial porosity before particle injection. We obtain $\phi(x,t)$ for the oil-free regions of the medium directly from the confocal micrographs using adaptive binarization, restricting all analysis to $\approx5$ bead diameters away from the inlet and outlet to minimize the influence of boundaries. Further, to reduce noise, we smooth the porosity measurements by segmenting the porous medium into representative elementary volumes (REV) $\approx10$ bead diameters in length, following common practice \cite{Blunt2017}. This level of smoothing provides a coarse-grained approach that results in a more realistic permeability estimate, but maintains enough fine-grained detail to incorporate any variations in deposition along the length of the medium. We thereby obtain $\varepsilon(x,t)$, which ranges from 0 to $\sim6$ \si{\micro\meter} over the course of the experiment, from \eqref{Porosity}.

Next, we use $\phi(x,t)$ and $\varepsilon(x,t)$ to estimate the permeability, $k(x,t)$, of the oil-free regions of the medium. Because $\varepsilon$ is more than an order of magnitude smaller than $a$, we do this by modifying the classic Kozeny-Carman model, which reasonably predicts the permeability of bead packings similar to those used in our experiment \cite{Garcia2009,Krummel2013}. Specifically, following typical convention \cite{Bear1972}, we model the pore space as a parallel bundle of cylindrical tubes of hydraulic radius equal to that of a packing of beads of radius $a+\varepsilon$. As detailed in \appref{Appendix A}, this model, along with \eqref{Porosity}, yields the relation
\begin{equation}
    k(x,t) = \frac{a^{2}}{18\tau}\left(\frac{1-\phi(x,t)}{1-\phi_{0}}\right)^{2/3}\frac{\phi(x,t)^3}{\left(1-\phi(x,t)\right)^2}
    \label{Kozeny Carman}
\end{equation}
where $\tau$ is the tortuosity of the pore space, assumed to be equal to 2 for simplicity \cite{Datta2013}. This relation quantifies the intuition that as deposition progresses and $\phi$ decreases, $k$ decreases as well. Thus, from our measurements of $\phi(x,t)$, we determine $k(x,t)$ \textit{via} \eqref{Kozeny Carman}.

\section{Model for Deposition-Induced Mobilization}
\begin{figure}[htp!]
\centering
\includegraphics[width=0.7\textwidth]{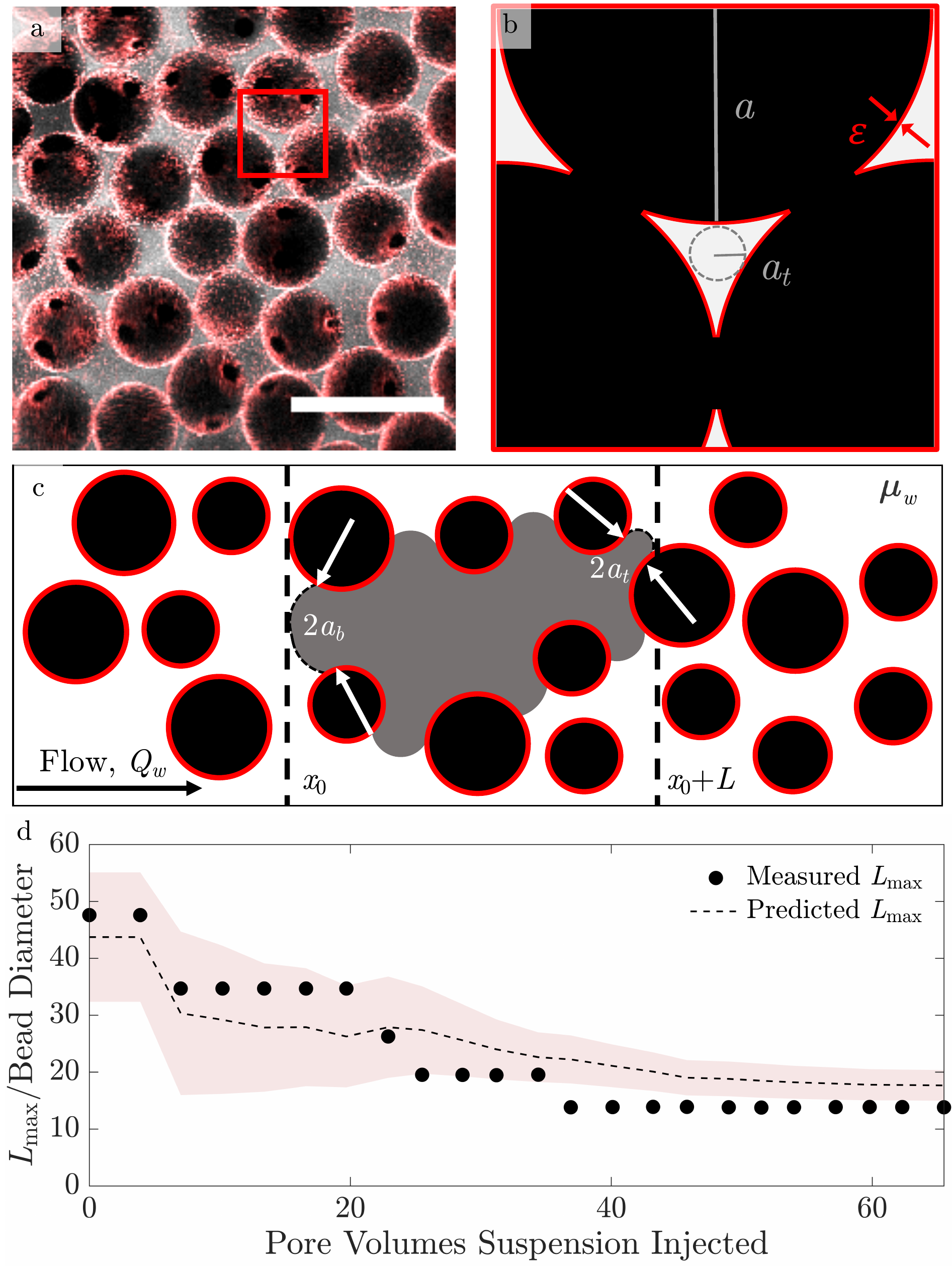}
\caption{Permeability reduction due to colloidal deposition promotes oil mobilization. (a) Magnified view of a confocal micrograph of colloidal deposition. White regions show pore space, black circles show the bead matrix, additional black regions show trapped oil ganglia, with the additional red showing deposited colloidal particles. Small black circles in the beads show the particle-free contacts between adjacent beads. Scale bar represents 250 \si{\micro\meter}. Imposed flow direction is from left to right. Deposition is nearly uniform over each bead surface. (b) Schematic of a pore throat of radius $a_t$, showing a uniform layer of deposited particles of thickness $\varepsilon$. (c) Schematic of a trapped oil ganglion. The grains that comprise the porous medium are shown black with red outlines to indicate deposition. The pore body and pore throat diameters, $2a_b$ and $2a_t$, are indicated at the upstream and downstream ends of the trapped ganglion at the positions $x_0$ and $x_0+L$ along the flow direction, respectively. We also indicate the wetting fluid flow rate, $Q_w$, with directionality from left to right and a constant viscosity, $\mu_w$.(d) Confocal measurements of the maximal length along the flow direction of trapped ganglia, $L_{max}$, over the course of colloidal injection. The initial point corresponds to Fig. 1d. As deposition progresses, increasingly smaller ganglia are mobilized from the medium, indicated by the decrease in the measured $L_{max}$. Dashed line shows our theoretical prediction of $L_{max}$ from Eqs. \ref{Lmax Criterion} and \ref{Lmax Simplified}. The red region shows uncertainty in the prediction arising from heterogeneities in deposition, calculated from $\pm$ the standard deviation in the measured $\phi$ over the medium.} 
\label{Figure 4}
\end{figure}

\noindent How does a deposition-induced reduction in permeability promote oil mobilization? To answer this question, we examine the competition between the viscous stress exerted by the flowing wetting fluid on a ganglion and the capillary pressure threshold that must be overcome to displace the ganglion. The gradient in the wetting fluid pressure $P_{v}$ is given by Darcy's law: $\frac{\partial P_{v}}{\partial x}=\frac{\mu_{w}\left(Q_{w}/A\right)}{\kappa k(x,t)}$, where the relative permeability $\kappa$ is an empirical parameter $\leq1$ used to quantify the modified transport through the medium due to the presence of trapped oil \cite{Blunt2017}. The viscous pressure drop $\Delta P_{v}$ across a ganglion of length $L$ along the imposed flow direction and with upstream end located a distance $x_{0}$ from the inlet can then be approximated as
\begin{equation}
         \Delta P_v \approx \frac{\mu_{w}\left(Q_{w}/A\right)}{\kappa}\int_{x_0}^{x_0+L}\frac{\mathrm{d}x}{k(x,t)}
    \label{Viscous Pressure Drop}
\end{equation}
where the ganglion spans $x_{0}\leq x\leq x_{0}+L$ as schematized in \figref{Figure 4}c. To mobilize the ganglion, this viscous pressure drop must be large enough to squeeze the ganglion through the pores of the medium. Specifically, the ganglion must displace the wetting fluid from a downstream pore, requiring it to overcome the capillary pressure threshold $2\gamma\cos\theta/a_{t}$, where $a_t$ is the radius of the downstream pore constriction, known as a ``throat", schematized in \figref{Figure 4}b. The ganglion must also be displaced by the wetting fluid from an upstream pore, requiring the capillary pressure within the pore to fall below the threshold $2\gamma\cos\theta/a_{b}$, where $a_b$ is the radius of the upstream pore, known as a ``body". Thus, to mobilize a ganglion from the medium, $\Delta P_{v}$ must exceed the capillary pressure threshold 
\begin{equation}
 \Delta P_c \approx 2\gamma\cos{\theta}\left[\frac{1}{a_t(x_0,t)}-\frac{1}{a_b(x_0+L,t)}\right].
    \label{Capillary Pressure}
\end{equation}
In previous studies, increasing the viscous stresses on ganglia was accomplished by increasing the wetting fluid flow rate, $Q_w$, to promote mobilization \cite{Datta2014}. Here, we examine another way of increasing these viscous stresses: by depositing particles in the pore space and thereby locally reducing permeability.

In the particle-free case, $k$, $a_{t}$, and $a_{b}$ are independent of position and time; \eqref{Viscous Pressure Drop} then reduces to $\Delta P_{v}\approx \mu_{w}\left(Q_{w}/A\right)L/\left(\kappa k_{0}\right)$, with the pristine medium permeability $k_{0}$ given by substituting $\phi=\phi_{0}$ into \eqref{Kozeny Carman}, and \eqref{Capillary Pressure} reduces to $\Delta P_{c}\approx 2\gamma\cos{\theta}\left(1/a_{t}-1/a_{b}\right)$. As a simplifying first step---motivated by the fact that our bead packings are comprised of beads with a narrow distribution of sizes---we adopt a mean-field approach for the pristine medium where $a_t$ and $a_b$ are averaged with $a_t\approx0.16a$ and $a_b\approx0.24a$ \cite{Al-Raoush2005,Thompson2008} instead of explicitly considering the pore-to-pore variation of the pore geometry. Ganglia are only mobilized if $\Delta P_{v}\geq\Delta P_{c}$, or equivalently, if their normalized length $\tilde{L}$ exceeds a threshold value
\begin{equation}
 \tilde{L}_{max} = \frac{\kappa k \cos{\theta}}{a\cdot\mathrm{Ca}}\left(\frac{1}{a_t}-\frac{1}{a_b}\right);
\end{equation}
here, overtildes indicate ganglion lengths that have been normalized by the bead diameter $2a$ and the Capillary number, defined as $\text{Ca}\equiv\mu_w\left(Q_w/A\right)/\gamma$, quantifies the competition between viscous and capillary stresses at the pore scale. This definition is based on the typical convention established in previous studies of flow through porous media \cite{Chatzis1983,Morrow1988}; it represents a global quantity defined using macroscopically-imposed injection conditions, and in our experiment, $\mathrm{Ca}\approx4\times10^{-6}$ is held constant. Previous permeability measurements performed in a similar porous medium using nearly identical fluids and flow conditions indicate that $\kappa\ll0.1$ for the range of residual oil saturations explored here \cite{Datta2014}. Furthermore, other relative permeability measurements  suggest that $\kappa$ varies only minimally in this range of oil saturations \cite{Wygal1963,Bartley2001,Ashrafi2012,Ghanbarian2017}. Therefore, for simplicity, we approximate the relative permeability as $\kappa\approx0.02$---a value that is both consistent with these previous findings and yields the best fit to our data. Using these experimental values, we expect that before particle injection, all ganglia longer than $\tilde{L}_{max}\approx44$ are mobilized---eventually establishing the initial steady-state configuration of trapped oil shown for $\approx6$--$50$ PVs in \figref{Figure 1}c and exemplified by the micrograph in \figref{Figure 1}d. Direct analysis of the confocal micrographs yields excellent agreement with this expectation: we find $\tilde{L}_{max}=48$ for this steady-state configuration in our experiment, as shown by the first data point in \figref{Figure 4}d, with all larger ganglia being mobilized in the first $\approx6$ PVs of wetting fluid injection. Thus, oil mobilization in a particle-free medium can be quantitatively described through our balance of viscous and capillary stresses.

After particle injection is initiated, $k$, $a_{t}$, and $a_{b}$ are no longer constant; $k(x,t)$ is given by \eqref{Kozeny Carman}, while $a_t(x,t) \approx 0.16a-\varepsilon(x,t)$ and $a_b(x,t) \approx 0.24a-\varepsilon(x,t)$, with $\varepsilon(x,t)$ given by \eqref{Porosity}. In this case, ganglia are again only mobilized if $\Delta P_{v}\geq\Delta P_{c}$, but now with $\Delta P_{v}$ and $\Delta P_{c}$ given by Eqs. \ref{Kozeny Carman} and \ref{Capillary Pressure}, respectively. Thus, we expect that ganglia are mobilized when their length $L$ exceeds a threshold value $L_{max}$ given by the solution to the following integral equation:
\begin{equation}
    \int_{x_0}^{x_0+L_{max}} \frac{\mathrm{d}x}{k(x,t)} = \frac{2\kappa\cos{\theta}}{\mathrm{Ca}}\left[\frac{1}{a_t\left(x_0+L_{max},t\right)}-\frac{1}{a_b\left(x_0,t\right)}\right].
    \label{Lmax Criterion}
\end{equation}
This equation quantifies the intuition that as colloidal deposition progresses, permeability reduction in the medium increases the viscous stresses on ganglia, decreasing $L_{max}$ and enabling increasingly smaller ganglia to be mobilized from the medium.

We again use our experimental measurements to test this prediction. Because particle deposition is nearly uniform throughout the medium, as shown in Figs. \ref{Figure 1}, \ref{Figure 2}, and \ref{Figure 3}, we average the values of $k(x,t)$ and $\varepsilon(x,t)$ predicted by Eqs. \ref{Kozeny Carman} and \ref{Porosity}, respectively, over the entire medium; the averaged values are given by $\bar{k}(t)\equiv\ell^{-1}\int_{0}^{\ell}k(x,t)\mathrm{d}x$ and $\bar{\varepsilon}(t)\equiv\ell^{-1}\int_{0}^{\ell}\varepsilon(x,t)\mathrm{d}x$, respectively. Substituting these averages in \eqref{Lmax Criterion} yields a more simplified prediction for the maximal length of a trapped ganglion: 
\begin{equation}
    \tilde{L}_{max}(t) = \frac{\kappa \bar{k}(t) \cos{\theta}}{a\cdot\mathrm{Ca}}\left(\frac{1}{\bar{a}_t(t)}-\frac{1}{\bar{a}_b(t)}\right),
    \label{Lmax Simplified}
\end{equation}
where $\bar{a}_t(t) \approx 0.16a-\bar{\varepsilon}(t)$ and $\bar{a}_b(t) \approx 0.24a-\bar{\varepsilon}(t)$. As before, $\mathrm{Ca}\approx4\times10^{-6}$ and we choose $\kappa = 0.02$, based on previous measurements obtained for a similar experimental system \cite{Datta2014}, and consistent with other studies \cite{Wygal1963,Bartley2001,Ashrafi2012,Ghanbarian2017}. Therefore, substituting the experimental values into \eqref{Lmax Simplified}, we expect that as particles are injected into the porous medium, all ganglia longer than $\tilde{L}_{max}$ are mobilized, with $\tilde{L}_{max}$ decreasing from $\approx44$ to $\approx18$ as deposition progresses,  indicated by the dashed line in \figref{Figure 4}d. That is, with increasing amounts of colloidal deposition, we expect that increasingly smaller ganglia can be mobilized from the medium. Direct analysis of the confocal micrographs again yields excellent agreement with this expectation: remarkably, we find that  $\tilde{L}_{max}$ decreases from $48$ to $15$ over the course of the experiment, as shown by the data points in \figref{Figure 4}d, reaching the final steady state after $\approx40$ PVs of suspension are injected. Thus, the configurations of ganglia mobilized through colloidal deposition can be quantitatively described through our balance of viscous and capillary stresses.

The approach to steady state in \figref{Figure 4}d is strikingly similar to that shown by the black points in \figref{Figure 1}c. Indeed, as a final test of our theory, we predict not just the maximal length of trapped ganglia, but the total amount of trapped oil, $S_{OR}$. We do this using our confocal micrographs by summing the pore space area occupied by all ganglia with lengths $\tilde{L}$ less than the $\tilde{L}_{max}$ value predicted by \eqref{Lmax Simplified}, at each experimental time point. The predicted $S_{OR}$ is shown by the dashed line in \figref{Figure 1}c. We again observe excellent agreement between the measured and predicted $S_{OR}$; both decrease over the course of colloidal injection and deposition, ultimately leading to $S_{OR}\approx20\%$ of its maximal value. Thus, the extent of deposition-induced oil mobilization can also be quantitatively described through our balance of viscous and capillary stresses.

\section{Discussion}
\noindent Our work reveals that particle deposition can promote mobilization of trapped oil ganglia from a porous medium. In particular, by reducing the local permeability, particle deposition increases the viscous stresses on ganglia, enabling them to overcome the influence of capillary stresses that keep them trapped. To identify the essential physics of this problem, our theory adopts a mean-field representation of the medium as a simplifying first step. Nevertheless, it yields predictions for the size and amount of mobilized fluid. Both of these quantities are in good agreement with the experimental measurements; however, incorporating spatial variations in pore geometry will be a useful next step that builds on the current work. 

Furthermore, our analysis only describes the \textit{onset} of ganglion mobilization, but does not consider subsequent ganglion dynamics. For example, we assume that as ganglia are mobilized, they do not interact with other ganglia as they traverse the porous medium, though coalescence is possible. As a result, ganglia that are smaller than the threshold predicted by our theory could be mobilized, potentially explaining the slight mismatch between our predicted values and measurements of the residual oil saturation at later times: specifically, our theory slightly over-predicts the measurements shown by the black points for $\sim$80--100 injected PVs in \figref{Figure 1}c. Additionally, it is possible that despite the strong attractive interactions between particles and glass beads, mobilized ganglia can remove some deposited particles \textit{via} capillary forces. This removal would result in a temporary increase in local permeability until particles redeposit there; we then expect that after sufficient re-deposition of particles, new ganglia can be mobilized, potentially enabling this process to repeat itself. This cyclic process could underlie the fluctuations in the overall amount of deposition measured at early times in our experiment, shown by the slight oscillations in the grey points for $\sim$60--80 injected PVs in \figref{Figure 1}c. It could also explain the slight mismatch between the predicted values and experimental measurements of the residual oil saturation at these times, shown by the dashed line and black points, respectively: because removal of particles through ganglion mobilization would result in a temporary increase in local permeability until particles redeposit there, mobilization of subsequent ganglia could be under-predicted by our theory, or could take longer than we predict. Elucidating these complex dynamics will be an important direction for future work.

Finally, we note that while our work has focused on the flow rate-controlled case, our results can also be modified to consider pressure-controlled flow. Unlike the flow rate-controlled case, uniform particle deposition under a constant pressure drop would not result in enhanced mobilization: the flow rate would necessarily decrease to compensate for the decreased permeability, maintaining a constant pressure drop across the medium, a constant pressure gradient along the medium, and therefore, an unchanged viscous pressure drop across each trapped ganglion. However, even for the case of a constant imposed pressure drop, particle deposition may still induce mobilization of trapped oil in the case of non-uniform (\textit{i.e.}, localized) deposition. Our theory enables us to establish a similar criterion to that presented in \eqref{Lmax Criterion} to predict mobilization under these circumstances, as detailed in \appref{Appendix B}.

\section{Conclusion}

\noindent Colloidal deposition in porous media is often considered to be a critical problem or, at best, a nuisance for applications including groundwater remediation and enhanced oil recovery; the permeability reduction resulting from deposition is thought to detrimentally impede subsequent flow and particle transport. Our work reveals that colloidal deposition can in fact be harnessed to mobilize and remove trapped immiscible fluids from a porous medium. In particular, we find that through deposition, a dilute ($<10^{-2}$ vol\%) suspension of particles can mobilize and remove an additional $\sim70\%$ of trapped fluid from a porous medium. Pore-scale visualization demonstrates that this mechanism of mobilization does not require colloidal surface activity or selective clogging of pores; instead, by reducing the local permeability, deposited particles increase the viscous stresses exerted on trapped fluid droplets by the surrounding fluid, enabling them to become mobilized. By analyzing the colloidal deposition profile and the pore-scale fluid stresses, we develop a geometric model that predicts which fluid droplets become mobilized, depending on their size (\eqref{Lmax Criterion})---in excellent agreement with our experimental results. Further, for a given starting distribution of trapped droplet sizes, this model enables us to predict the extent of fluid that is mobilized as deposition progresses---again in excellent agreement with our experimental results. The model can also be used to guide applications requiring a target amount of trapped fluid to be mobilized: given a desired value of $S_{OR}$, Eqs. \ref{Kozeny Carman} and \ref{Lmax Criterion} can be inverted to determine the extent of colloidal deposition required. Thus, in addition to shedding light on a new way by which colloids can mobilize trapped fluid from a porous medium, our work also provides quantitative guidelines for applying this mechanism. Because it does not require specialized physico-chemical characteristics, we expect this mechanism could be utilized in diverse media of different structures, and applicable to colloids of differing chemistry and physical characteristics, including naturally-occurring organic and inorganic colloids e.g., ``fines".

% Acknowledgements and Author Contributions
\begin{acknowledgments}
\noindent 
It is a pleasure to acknowledge C.A. Browne, J.A. Ott, and N. Bizmark for helpful feedback on the manuscript, as well as H.A. Stone and I.C. Bourg for stimulating discussions. This work was supported by the Grand Challenges Program of the High Meadows Environmental Institute. J.S. was also supported in part by the Mary and Randall Hack Graduate Award of the High Meadows Environmental Institute. 
\end{acknowledgments}

% Placeholder for author contributions
\authorcontributions{J.S. and S.S.D. designed the experiments; J.S. performed all experiments; J.S. and S.S.D. analyzed the data and developed/implemented the theoretical model; S.S.D. designed and supervised the overall project. All authors discussed the results and wrote the manuscript.}

\newpage% Placeholder for appendices
\begin{appendices}

\numberwithin{equation}{section}
\section{Model for permeability alteration by deposition}
\label{Appendix A}
\noindent We model the pore space as a parallel bundle of cylindrical tubes of radius $R$ and length $\ell'>\ell$, where $\left(\ell'/\ell\right)^{2}$ is defined as the hydrodynamic tortuosity $\tau$. The mean flow speed in each tube, $\langle v \rangle$, is then given by the Hagen-Poiseuille equation:
\begin{equation}
    \langle v \rangle \equiv \frac{Q}{\phi A} = \frac{R^2 \Delta P}{8\mu_w \ell'}
    \label{Tube Velocity}
\end{equation}
where $Q$ is the imposed volumetric flow rate through the entire medium, $A$ is the cross-sectional area of the mediun, $\phi$ is the medium porosity, and $\Delta P$ is the pressure drop across the entire medium. This mean flow speed is also given by Darcy's law: 
\begin{equation}
    \langle v \rangle 
    \equiv \frac{Q}{\phi A} = \frac{k\Delta P}{\phi\mu_{w}\ell}
    \label{Pore Velocity}
\end{equation}
where $k$ is the overall permeability of the medium. Equating Eqs. \ref{Tube Velocity} and \ref{Pore Velocity} thus yields a general expression for the permeability:
\begin{equation}
    k = \frac{\phi R^2}{8\tau}.
    \label{General k}
\end{equation}
Following typical convention \cite{Bear1972}, we solve for $R$ by equating the hydraulic radii $R_{h}$ of the tube bundle and the porous medium composed of beads of radius $a+\varepsilon$. For the tube bundle, this quantity is given by the ratio between the pore space cross-section and the wetted perimeter, $R_h = \frac{\pi R^2}{2\pi R}=R/2$. For the porous medium, this quantity is given by the ratio between the pore space volume and the wetted surface area, $R_h = \frac{\phi A\ell}{N\times4\pi\left(a+\varepsilon\right)^2}$, where $N=\frac{\left(1-\phi\right)A\ell}{\frac{4}{3}\pi\left(a+\varepsilon\right)^3}$ is the number of beads in the medium. Thus, equating both expressions for $R_{h}$ yields
\begin{equation}
    R = \frac{2\phi\left(a+\varepsilon\right)}{3\left(1-\phi\right)}
    \label{R Final}
\end{equation}
Finally, substituting \eqref{R Final} into \eqref{General k}, and applying \eqref{Porosity} from the main text, yields our generalized prediction of the permeability, \eqref{Kozeny Carman}.

To highlight the impact of permeability reduction on enhanced mobilization, we contrast a porous medium in which particle deposition alters the porosity to that of a medium in which the pores have simply become uniformly smaller. If we consider the case of a pristine porous medium with smaller pores, the viscous pressure drop across a trapped ganglion of length $L$, given by $\Delta P_v \propto L/k$, is larger because $k$ scales as $a^2$, where $a$ is the bead radius. However, it is important to note here that the capillary pressure threshold that must be overcome to mobilize the ganglion, given by $\Delta P_c \propto 1/a$, is also larger in a medium with smaller pores. Balancing the two confirms that the normalized threshold ganglion length, $L_{max}/2a$, is a constant and is thus unchanged when pore size is reduced, as one may expect. However, in the case of a porous medium in which particle deposition has altered both the effective grain size $a$ and the porosity $\phi$, the permeability decreases more than the capillary pressure threshold due to its dependence on porosity. The capillary pressure threshold increases as $1/(a+\varepsilon)$; however, this increase is not sufficient to overcome the increase in the viscous pressure drop due to the decrease in permeability, which decreases as $(a+\varepsilon)^2 \phi^3/(1-\phi)^2$. Specifically, $\Delta P_v \propto \frac{L}{(a+\varepsilon)^2}\frac{(1-\phi)^2}{\phi^3}$ and $\Delta P_c \propto a/(a+\varepsilon)$; balancing the two in this case yields a normalized threshold ganglion length $\frac{L_{max}}{2(a+\varepsilon)}\propto \frac{\phi^3}{(1-\phi)^2}$ that is not a constant, but also reduces when porosity is reduced.

\newpage\section{Theory for pressure-controlled flow} \label{Appendix B}
\begin{wrapfigure}{l}{0.5\textwidth}
    \centering
    \includegraphics[width=0.5\textwidth]{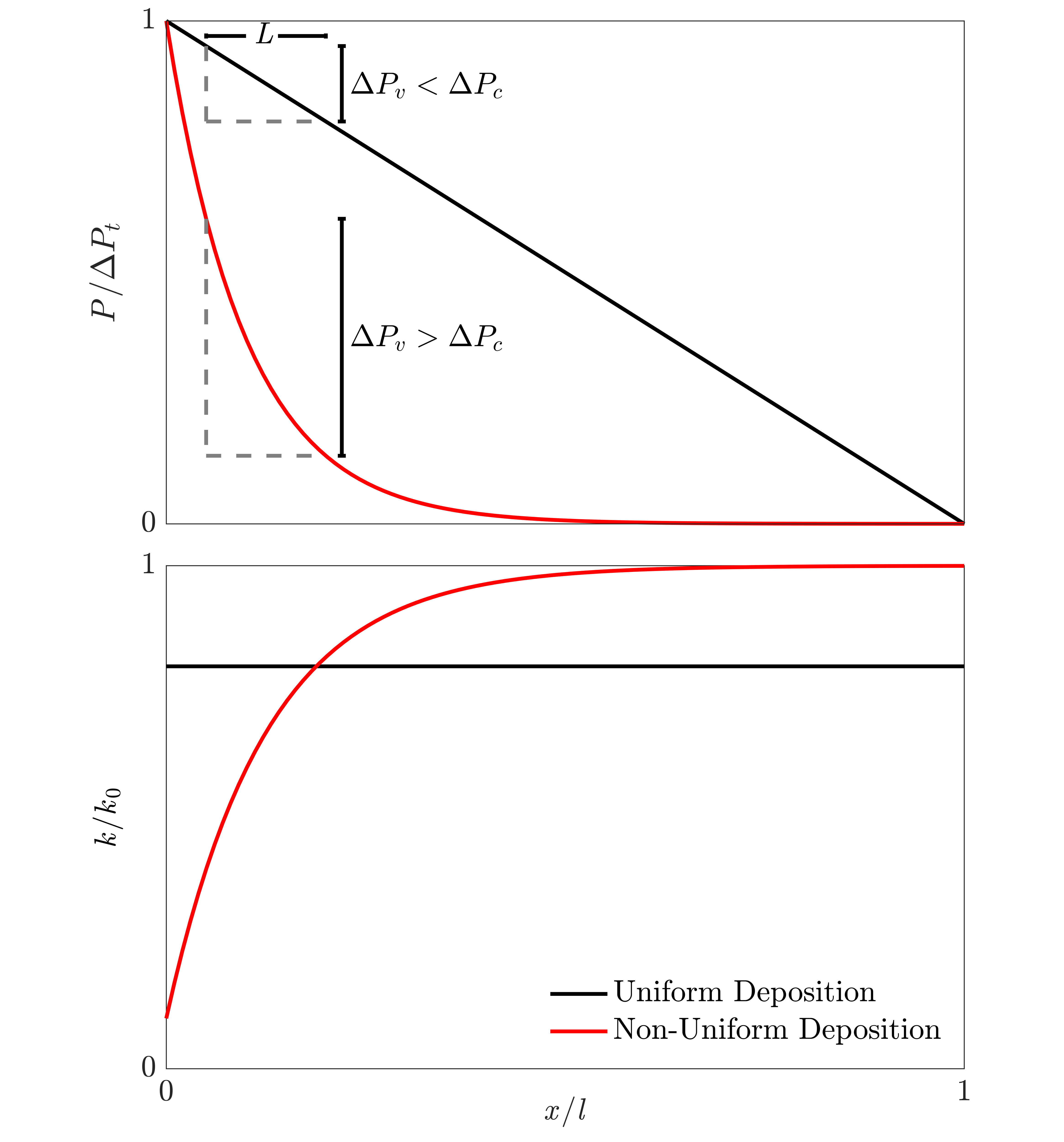}
    \caption{Non-uniform deposition promotes mobilization under constant pressure drop $\Delta P_t$. (top) Fluid pressure normalized by the pressure drop $\Delta P_t$. A ganglion of length $L$ has $\Delta P_v < \Delta P_c$ when particle deposition is spatially uniform at a low fixed $\Delta P_t$ (black). However, a non-uniform deposition profile obtained at the same constant $\Delta P_t$ can result in an increase of $\Delta P_v > \Delta P_c$ (red). (bottom) The corresponding profiles of normalized permeability, $k/k_0$, are shown over the length of the porous medium, $x/\ell$.}
    \label{Constant Pressure}
\end{wrapfigure}

\noindent To modify our theory for the case of a constant imposed pressure drop, $\Delta P$, we consider both uniform and non-uniform deposition of particles along the flow direction. Though we do not expect to see enhanced mobilization in the case of uniform particle deposition under a constant imposed pressure drop, particle deposition may still induce mobilization in the case of non-uniform (\textit{i.e.}, localized) deposition. An example is sketched in \figref{Constant Pressure}. Here, we contrast a uniform deposition profile (black curves) under a constant total pressure drop across the medium, $\Delta P_t$, with a non-uniform deposition profile in which more particles are deposited near the inlet of the medium (red curves). The corresponding permeability curves, $k/k_0$, where $k_0$ is the permeability of the pristine medium, are shown in the lower panel of the figure; uniform deposition corresponds to a uniform $k$ along the medium (black curve), while non-uniform deposition results in lower $k$ near the inlet (red curve), where $x$ is the position along the medium and $\ell$ is its length. The resultant variation of the fluid pressure $P/\Delta P_t$ along the porous medium is shown by the plot in the top panel. 

In the case of uniform deposition at constant $\Delta P_t$, the flow rate $Q$ decreases concomitantly with increasing deposition. As a result, the viscous stress over a ganglion would not vary considerably, and thus new ganglia are not likely to be mobilized once deposition begins. An example of a ganglion of length $L$ is shown by the horizontal dashed line in the top panel of the figure; the vertical bar indicates the viscous pressure drop across it, $\Delta P_v$, which in this case is not sufficient to exceed the capillary pressure drop $\Delta P_c$, and it remains trapped. As uniform deposition progresses, this viscous pressure drop does not change, and the ganglion still remains trapped. However, in the case of a non-uniform deposition profile, due to the non-uniform pressure gradient along the medium, the viscous pressure drop along the same ganglion is much larger and exceeds $\Delta P_c$, as indicated by the lower vertical bar. In this case, the ganglion becomes mobilized. Thus, even in pressure-controlled flow, particle deposition can induce mobilization when it is non-uniform.

While our experiments focus on the flow-controlled case, our theoretical analysis can be adopted to model the case of mobilization under pressure-controlled flow due to non-uniform deposition. Specifically, \eqref{Viscous Pressure Drop} of our manuscript would then be modified to be:
\begin{equation}
    \Delta P_v = \frac{\mu_w}{A}\int_{x_0}^{x_0+L} \frac{Q_w(x,t)}{k'(x,t)}\mathrm{d}x
    \label{Modified Viscous Pressure Drop}
\end{equation}
where $k'(x,t)=\kappa(x,t)k(x,t)$. Here, we have two unknowns, $Q$ and $k'$. However, we also have another relationship that comes from the overall constant pressure drop criterion as follows:
\begin{equation}
\Delta P_t = \mathrm{const.} = \frac{\mu_w}{A}\int_{0}^{\ell} \frac{Q_w(x,t)}{k'(x,t)}\mathrm{d}x
\label{Total Pressure Equation}
\end{equation}
Using these modified equations in conjunction with \eqref{Capillary Pressure} in the main text of the manuscript would then provide a modified criterion for ganglion mobilization \textit{via} particle deposition under a constant imposed pressure drop.

\end{appendices}

\cleardoublepage

\end{document}